\documentclass[aip,graphicx,preprint]{revtex4-1}
\usepackage{graphicx}
\usepackage[caption=false]{subfig}
\usepackage{xcolor}

\draft

\colorlet{blue}{black}

\begin{document}
	
	\title{Amplitude of jump motion signatures in classical vibration-jump dynamics}
	
	\author{Peter S. M. Townsend}
	\email[]{psmt2@alumni.cam.ac.uk}
	\altaffiliation[Present address: ]{Rutgers University Department of Chemistry and Chemical Biology, 123 Bevier Road, Piscataway Township, NJ 08901}
	\affiliation{Cavendish Laboratory, J.J. Thomson Avenue, Cambridge CB3 0HE, United Kingdom}
	
	\author{John Ellis}
	\affiliation{Cavendish Laboratory, J.J. Thomson Avenue, Cambridge CB3 0HE, United Kingdom}
	
	\date{\today}
	
	\begin{abstract}
		The classical Langevin dynamics of a particle in a periodic potential energy landscape are studied via the intermediate scattering function (ISF). By construction, the particle performs coupled vibrational and activated jump motion with a wide separation of the vibrational period and the mean residence time between jumps. The long time limit of the ISF is a decaying tail proportional to the function that describes ideal jump motion in the absence of vibrations. The amplitude of the tail is unity in idealized jump dynamics models, but is reduced from unity by the intra-well motion. Analytical estimates of the  amplitude of the jump motion signature are provided by assuming a \textcolor{blue}{factorization of the conditional probability density of the particle position at long times, motivated by the separation of time scales associated with inter-cell and intra-cell motion}. The assumption leads to a factorization of the ISF at long correlation times, where one factor is an ideal jump motion signature, and the other component is the amplitude of the signature. The amplitude takes the form of a single-particle anharmonic Debye-Waller factor. The factorization approximation is exact at the diffraction conditions associated with the periodic potential. Numerical simulations of the Langevin equation in one and two spatial dimensions confirm that for a strongly corrugated potential the analytical approximation provides a good qualitative description of the trend in the jump signature amplitude, between the points where the factorization is exact. Published full-text \url{https://doi.org/10.1063/1.5053123}
	\end{abstract}
	
	\maketitle
	
	\section{Introduction}
	\label{sec:intro}
	
	Surface diffusion at the atomic scale is a unifying area of study linking a fundamental theoretical and experimental understanding of model systems \cite{AlaNissila2002AdvPhys,MiretArtes2005JPCM,Jardine2009ProgSurfSci} to diverse applications including the self-assembly of carbon sheets \cite{Hofmann2005PRL,Sutter2008NatMat}, ordered organic \cite{Love2005ChemRev} and biological \cite{Sun2009JACS} molecular layers, and quantum dots \cite{Petroff2001PhysToday}. The mobility of adsorbed species depends critically on the interplay between vibrations and over-barrier hopping, driven by fluctuation and dissipation \cite{MiretArtes2005JPCM,Jardine2007JPCM,Lechner2015PCCP}. Both vibrational and diffusive motion are sensitive measures of the mean adsorbate/surface interaction potential and also the strength of dissipative effects. Experimentally, vibrational and diffusive surface dynamics can be studied on equal footing using energy resolved scattering techniques \cite{Toennies1993JPCM,Graham2003SurfSciRep,Fouquet2005RSI,Jardine2009ProgSurfSci}. The present work addresses the relative intensities of different regions of the inelastic scattering spectrum in such experiments, as predicted by simple kinematic scattering and dissipative molecular dynamics.
	
	To interpret the results of scattering experiments on mobile overlayers, the effect of the substrate heat bath is commonly modeled by Langevin dynamics \cite{Graham1997PRB,Hedgeland2009PRB,Hedgeland2011PRL,Lechner2013JCP,Rotter2016NatMat}, which can be used to combine theoretical and experimental inputs to learn about the microscopic origins of atomic-scale dissipation \cite{Rittmeyer2016PRL}. A natural and standard way to describe statistical motion such as trajectories in Langevin dynamics, is via conditional probabilities or equivalently by correlation functions \cite{Zwanzig1964AnnuRevPhysChem,Harp1970PRA}. The intermediate scattering function (ISF) is one such correlation function and provides a comprehensive mathematical description of different types of the equilibrium dynamics \cite{Jardine2009ProgSurfSci}. Additionally, the ISF for an ensemble of diffusing adsorbed species can be probed experimentally by the helium-3 surface spin echo method (HeSE) \cite{Jardine2009ProgSurfSci}, at length and time scales where the results are directly sensitive to the detailed nature of the potential energy landscape and the dissipation mechanism \cite{Ward2013Thesis}. 
	
	Results relating to absolute quasi-elastic scattering intensities in the HeSE experiment have an important contribution to make towards the interpretation of the underlying surface dynamics. Up to now most HeSE experiments have focused on the slowly decaying tail of the ISF and its interpretation in terms of jump motion. However, the fast decay of the ISF at short correlation times contains a wealth of information on the short time scale dynamics. Under certain circumstances the fast decay can provide a direct measure of the friction \cite{Lechner2015PCCP}, and for weakly corrugated adsorption systems the fast decay dominates the experimental signature, allowing continuous diffusive \cite{Hedgeland2009NatPhy} and ballistic \cite{Ellis1999PRL} motion to be resolved. In the opposite limit of strongly corrugated lateral potentials, the ISF separates into fast and slow components, whose amplitudes are constrained by the absolute normalization of the correlation function. The amplitude of the component describing jump diffusion is typically substantially less than unity in experiments \cite{Fouquet2006PhysicaB,Jardine2007JPCM,Jardine2008NJP,Hedgeland2009PRB,Paterson2011PRL,Kole2012JPCM}, and the amplitude typically decreases with increasing surface-parallel momentum transfer of the scattering probe \cite{Kole2012JPCM,Ward2013Thesis}. The low amplitude may be due to a number of effects, including inelastic scattering from surface phonons \cite{Benedek1994SurfSci,Benedek2010JPCM,MartinezCasado2010JPCM}. However, the intra-cell dynamics of the adsorbates diffusing in a continuous potential also contribute to a reduction in the amplitude of the jump motion signature \cite{Jardine2009PCCP}.
	
	The line shapes describing pure jump motion on a Bravais lattice, completely neglecting intra-cell motion, are well known \cite{Chudley1961ProcRoySoc}. In the time domain they consist of mono-exponential decays with unit amplitude. In the opposite limit of confined diffusion with no jumps, analytical results are known for dephased harmonic vibrations \cite{Vega2004JPCM}. Previous attempts to synthesise the vibration and jump motion limits into a unified description of motion in a sinusoidal potential, have included the use of a cumulant expansion method combined with variable parameters that allow the amplitude of different components to be adjusted\textcolor{blue}{\cite{MartinezCasado2007JCP,MartinezCasado2007PRL}}. In the present work the amplitude of the jump diffusion component of the ISF is estimated from the continuous potential with no adjustable parameters. The amplitude depends only weakly on the strength of the Langevin friction, as long as the dynamics are cleanly separated into intra-well and jump motion. The amplitude is estimated without dynamical simulation using an approximate factorization of the ISF at long times into a decaying factor depending on the jump dynamics, and a constant factor depending on the intra-cell thermodynamics. In Section \ref{sec:analytical} the conceptual basis of the factorization approximation is established, and the resulting estimate for the amplitude of the jump component is shown to be an interpolation between exact \textcolor{blue}{static levels of the ISF} at the diffraction conditions \textcolor{blue}{of the periodic potential}. In Section \ref{sec:numerical}, the factorization approximation is shown to closely match the jump signature amplitudes fitted from numerical one-dimensional, and non-separable two-dimensional, Langevin simulations.
	
	\section{Vibration-jump diffusion}
	\label{sec:analytical}
	
	Consider the dynamics of a particle, mass $m$, obeying a one-dimensional classical Langevin Equation (LE), with friction $\gamma$, in a potential of mean force $V(x)$. The friction represents the decay rate of the velocity autocorrelation function in the absence of the periodic potential \cite{Berne1966JCP,MiretArtes2005JPCM}. The LE for the particle co-ordinate $x$ reads:
	\begin{equation}
	m\ddot{x}=-V'(x)-m\gamma \dot{x} + F(t)\,\textrm{,}
	\end{equation}
	in which $F(t)$ is a random force of zero mean and autocorrelation $\langle F(t)F(0) \rangle = m\gamma k_{B}T\delta(t)$ thereby satisfying the classical fluctuation-dissipation theorem \cite{Kubo1966RepProgPhys}. The (generalized) LE is known to impose thermal equilibrium on the particle \cite{Ottobre2011Nonlin}, meaning that the long-term distribution of the position is the canonical distribution,
	\begin{equation}
	\label{eqn:classicalEquilibriumDistribution}
	p(x)\propto e^{-\beta V(x)}\,\textrm{.}
	\end{equation}
	The ISF depends on time-dependent conditional probabilities which in general cannot be expressed so succinctly. However, the canonical distribution will retain a central importance when estimating the decay amplitude of the tail of the ISF representing jump motion.
	
	\subsection{Classical ISF}
	\label{sec:classicalISF}
	
	The classical ISF $I(\Delta K,t)$ for one dimensional motion is the spatial Fourier transform of van Hove's conditional probability function $G(x,t)$ \cite{VanHove1954PhysRev} or equivalently the autocorrelation of a kinematic scattering amplitude $e^{i\Delta K x(t)}$. Explicitly,
	\begin{equation}
	\label{eqn:genericClassicalISF}
	I(\Delta K,t)=\langle e^{i\Delta K x(t)}e^{-i\Delta K x(0)}\rangle\,\textrm{,}
	\end{equation}
	where the angle brackets represent an ensemble average, or equivalently a time average under ergodicity which is another provable feature of the LE \cite{Ottobre2011Nonlin}.
	
	It will prove useful to unpack the meaning of the average $\langle \cdots \rangle$ in Equation(\ref{eqn:genericClassicalISF}) in terms of explicit conditional probabilities. Let $p(x)$ represent the equilibrium (canonical) probability density of the particle residing at $x$, and \textcolor{blue}{$p_{t}(x'|x)$} the conditional probability density to move \textcolor{blue}{from $x$} to $x'$ in time $t$, averaged over any other co-ordinates and momenta in the system. Then, from the general definition of a classical autocorrelation function, the ISF can be written as:
	\textcolor{blue} 
	{\begin{equation}
		\label{eqn:isfDefOpen}
		I(\Delta K,t)=\int dx\,dx'\, p(x) e^{i\Delta K x'}e^{-i\Delta K x} p_{t}(x'|x)\,\textrm{.}
		\end{equation}}
	The representation (\ref{eqn:isfDefOpen}) makes no explicit reference to environmental degrees of freedom to which the particle is coupled, but defines the ISF in terms of a stochastic process characterized by $p(x)$ and \textcolor{blue}{$p_{t}(x'|x)$}, which are to be interpreted as being pre-averaged over all co-ordinates and momenta that affect the time evolution of the particle co-ordinate.
	
	\subsection{Harmonic systems}
	\label{sec:harmonic}
	
	It is convenient to review the ISF for a classical Langevin particle in harmonic potential, in order to motivate the intuitive approximation underlying the later derivation of decay amplitudes. In the harmonic potential
	\begin{equation}
	V(x)=\frac{1}{2}m\Omega^{2}x^{2}\,\textrm{,}
	\end{equation}
	the ISF has been previously \textcolor{blue} {\cite{Guantes2004JCP,MartinezCasado2007JCP,MartinezCasado2010ChemPhys}} derived as
	
	\begin{widetext}
		\textcolor{blue}
		{\begin{equation}
			\label{eqn:harmonicISF}
			I(\Delta K,t)=\exp\Big(-\frac{k_{B}T}{m\Omega^{2}}\Delta K^{2}\Big[1-e^{-\gamma t/2}\{\cos(ft)+\frac{\gamma}{2f}\sin(ft)\}\Big]\Big)\,\textrm{,}
			\end{equation}}
	\end{widetext}
	where $f=\sqrt{\Omega^{2}-\gamma^{2}/4}$. 
	
	The key feature of the analytical result (\ref{eqn:harmonicISF}), for the present argument, is the long time limit (static level) ${I(\Delta K,t\rightarrow\infty)}$:
	\begin{equation}
	\label{eqn:harmonicLongTimeLimit}
	I(\Delta K,t\rightarrow\infty)=\exp\Big(-\frac{k_{B}T}{m\Omega^{2}}\Delta K^{2}\Big)\,\textrm{.}
	\end{equation}
	The static level of the ISF is nonzero, as a result of the confined nature of the diffusion \cite{Jardine2009PCCP}. Additionally, the static level is a thermodynamic quantity independent of the dissipation strength.
	
	The static level can be computed directly, without computing the entire ISF, by setting $p_{t}(x'|x)=p(x')$ in the definition (\ref{eqn:isfDefOpen}). In other words we assume that due to the action of the heat bath with a finite friction, at long enough correlation times the particle position $x'$ follows its own independent canonical distribution regardless of the initial position. The ISF can then be computed as the product of two time-independent integrals:
	\begin{equation}
	\label{eqn:isfDefOpenFactorized}
	I(\Delta K,t\rightarrow\infty)=\Big(\int dx\, p(x) e^{i\Delta Kx}\Big)\Big(\int dx\, p(x) e^{-i\Delta Kx}\Big)
	\end{equation}
	\begin{equation}
	\label{eqn:classicalIntraCellFactorHarmonic}
	=\Bigg|\frac{\int dx\, e^{-\beta m\Omega^{2}x^{2}/2} e^{i\Delta Kx}}{\int dx' \, e^{-\beta m\Omega^{2}x'^{2}/2}}\Bigg|^{2}
	\end{equation}
	\begin{equation}
	=\exp\Big(-\frac{k_{B}T}{m\Omega^{2}}\Delta K^{2}\Big)\,\textrm{,}
	\end{equation}
	in agreement with Equation (\ref{eqn:harmonicISF}). A similar factorization argument will now be used to derive an approximate result giving the amplitude of the long-time decaying tail of the ISF when intra-cell vibrations are coupled to inter-cell jumps.
	
	\subsection{Factorization approximation}
	\label{sec:factorization}
	
	The ISF describing perfect jump diffusion with jumps of lengths $\{j\}$ and rates $\{\Gamma_{j}\}$ on a one-dimensional Bravais lattice of site separation $a$, with no intra-cell motion, is a mono-exponential decay with unit amplitude and $\Delta K$-dependent decay rate $\alpha(\Delta K)$ given by the Chudley-Elliott model \textcolor{blue}{\cite{Chudley1961ProcRoySoc}}:
	\textcolor{blue}
	{\begin{equation}
		I(\Delta K,t)=\exp[-\alpha(\Delta K),t]\,\textrm{;}
		\end{equation}}
	\begin{equation}
	\label{eqn:ceModel1d}
	\alpha(\Delta K)=2\sum_{j}\Gamma_{j}\sin^{2}\Big(\frac{\Delta K aj}{2}\Big) \textrm{.}
	\end{equation}
	In the analysis of motion in a continuous periodic potential, sufficiently strongly corrugated that mean jump rates are significantly slower than intra-cell time scales, we can therefore define a decay amplitude $A(\Delta K)$ by
	\begin{equation}
	\label{eqn:longTimeLimit}
	I(\Delta K,t>t^{*})\approx A(\Delta K)\exp\Big[-\alpha(\Delta K)t\Big]\,\textrm{,}
	\end{equation}
	where $t^{*}$ is a time beyond which the ISF is well described by an exponential decay. \textcolor{blue}{In other words, in the present work we assume that the ISF for all times can be written as
		\begin{equation}
		I(\Delta K,t)=A(\Delta K,t)\exp\Big[-\alpha(\Delta K)t\Big]
		\end{equation}}
	\textcolor{blue}{where $A(\Delta K,t)$ decays to a constant, $A(\Delta K)$, at long times.}
	
	At any given $\Delta K$, both $A$ and $\alpha$ can be determined from simulated ISFs via the best exponential fit to the long-time tail. Comparison of $\alpha(\Delta K)$ with the Chudley-Elliott model defines the jump distribution, and is a standard data reduction method used to compare molecular dynamics simulations and HeSE data \cite{Alexandrowicz2006PRL,Alexandrowicz2008JACS,Lechner2012Thesis,Ward2013Thesis}. Predicting $\alpha(\Delta K)$ is an application of rate theory \cite{Hanggi1990RevModPhys},\textcolor{blue}{\cite{Guantes2003JCP}}, whereas the present work focuses on developing an approximate theory for $A(\Delta K)$. Phrased in the frequency domain, it is well known that the relative amplitude of the diffusive and vibrational contributions to the dynamical structure factor are $\Delta K$-dependent \cite{Chen1994PRB}, with the vibrational (T-mode) contribution being an increasing function of $\Delta K$ \cite{Braun1997JCP,Graham1997PRB}. We now derive an approximate analytical expression for the decreasing function describing the amplitude remaining in the diffusive jump-motion signature.
	
	\textcolor{blue}{Write the particle co-ordinate $x(t)$ as the sum of an integer-valued co-ordinate $n(t)$ describing the unit cell number, and a residual intra-cell co-ordinate $0\leq w(t)<a$:}
	\begin{equation}
	x(t)=an(t)+w(t)\,\textrm{.}
	\end{equation}
	All integrals over $w$ and $w'$ will have implicit limits of $0$ and $a$. To avoid a proliferation of function names describing probabilities, the label $p()$ will be used to represent different probability functions depending on the number and type of arguments. $p(w)$ or $p(w')$ represent canonical probability density in a single unit cell. $p_{t}(\cdot | \cdot)$ represents the \textcolor{blue}{$t$-dependent} probability or probability density of everything to the left of the divide $|$ conditional \textcolor{blue}{on} everything to the right of the divide, regardless of the number of co-ordinates to the left or right of the divide. 
	
	The formal expansion (\ref{eqn:isfDefOpen}) of the ISF in terms of conditional probabilities then reads
	\begin{widetext}
		\begin{equation}
		\label{eqn:formalISFnwExpansion}
		I(\Delta K,t)=\int dw\,dw'\,\sum_{n,n'} p_{t}(n',w'|n,w)e^{i\Delta K(an'+w')}e^{-i\Delta K(an+w)}p(n,w)\,\textrm{,}
		\end{equation}
	\end{widetext}
	where $p(x)\rightarrow p(n,w)$ and $p_{t}(x'|x)\rightarrow p_{t}(n'w'|n,w)$ have been substituted. 
	
	\color{blue}
	
	Since all initial sites are equivalent, we can choose an arbitrary initial site, and we choose $n=0$, so the ISF (\ref{eqn:formalISFnwExpansion}) is given by
	\begin{widetext}
		\begin{equation}
		\label{eqn:formalISFnwExpansionZeroed}
		I(\Delta K,t)=\int dw\,dw'\,\sum_{n'} p_{t}(n',w'|0,w)e^{i\Delta K(an'+w')}e^{-i\Delta Kw}p(w)\,\textrm{,}
		\end{equation}
	\end{widetext}
	
	To make further progress with Equation (\ref{eqn:formalISFnwExpansion}), which contains all the full complexity of the nonlinear dissipative dynamics, we introduce a factorization approximation based on the separation of timescales in jump dynamics. We assume that as the conditional probability distribution $p_{t}(n',w|0,w)$ spreads out from the initial unit cell, the intra-cell relaxation is much faster than the mean jump rate. Then, at intermediate times we should expect $p_{t}(n',w'|0,w)$ to factorize as
	\begin{equation}
	\label{eqn:factorizationApprox}
	p_{t}(n',w'|0,w)=p(w')p_{t}(n'|0,w)\,\textrm{,}
	\end{equation}
	where $p(w')$ describes the same canonical equilibrium distribution as $p(w)$, and the remaining factor $p_{t}(n'|0,w)$ describes the probability of the particle being found in the unit cell of index $n'$. Substituting the factorized conditional probability into the ISF (\ref{eqn:formalISFnwExpansionZeroed}) gives
	\begin{widetext}
		\begin{equation}
		I(\Delta K,t)=\int dw\,dw'\,\sum_{n'} p(w') p_{t}(n'|0,w)e^{i\Delta K(an'+w')}e^{-i\Delta Kw}p(n,w)\,\textrm{,}
		\end{equation}
	\end{widetext}
	which can be written as
	\begin{equation}
	\label{eqn:ISFqExpansion}
	I(\Delta K,t)=\textcolor{blue}{\sum_{n'}}e^{i\Delta Kan'}q_{t}(n')\,\textrm{,}
	\end{equation}
	where the $q_{t}(n')$ are integrals over initial and final intra-cell co-ordinates:
	\begin{equation}
	\label{eqn:wIntegralqForm}
	q_{t}(n')=\int dw\, dw'\, p(w')e^{i\Delta Kw'}e^{-i\Delta Kw} p_{t}(n'|0,w)p(w)\,\textrm{.}
	\end{equation}
	
	We calculate the amplitude $A(\Delta K)$ in Equation (\ref{eqn:longTimeLimit}) by noting that the Chudley-Elliot result (\ref{eqn:ceModel1d}) can be expressed in terms of time-dependent site probabilities $P_{t}(n')$ as
	\begin{equation}
	I(\Delta K,t)=\sum_{n'} e^{i\Delta Kan'} P_{t}(n')\,\textrm{,}
	\end{equation}
	where the functions $P_{t}(n')$ satisfy
	\begin{equation}
	\label{eqn:unityNormalization}
	\sum_{n'} P_{t}(n')=1
	\end{equation}
	at all times. If the ISFs derived from Langevin dynamics simulations in a corrugated potential display as their long time limit mono-exponential tails whose $\alpha(\Delta K)$ agrees with a Chudley-Elliott model, then at long times the functions $q_{t}(n')$ in Equation (\ref{eqn:ISFqExpansion}) stay in the correct proportions, as they evolve, to describe jump motion. Then, at long times the functions $q_{t}(n')$ are proportional to some set of probabilities $P_{t}(n')$ that are normalized to unity as described by Equation (\ref{eqn:unityNormalization}). The amplitude of the exponential decay in the Chudley-Elliott model is also unity, and therefore the exponential decay amplitude implied by Equation (\ref{eqn:ISFqExpansion}) is the normalization of the set of $q_{t}(n')$ functions:
	\begin{equation}
	A(\Delta K)=\lim_{t\rightarrow \infty}\sum_{n'}q_{t}(n')\,\textrm{.}
	\end{equation}
	Substituting (\ref{eqn:wIntegralqForm}), and bringing the $n'$-sum inside the $w,w'$-integrations, the amplitude is expressed as
	\begin{equation}
	\label{eqn:amplitudeFormal}
	A(\Delta K)=\int dw\, dw'\, p(w')e^{i\Delta K(w'-w)}p(w) \sum_{n'} p_{t}(n'|0,w)\,\textrm{.}
	\end{equation}
	However, by construction
	\begin{equation}
	\sum_{n'} p_{t}(n'|0,w)=1\,\textrm{,}
	\end{equation}
	because the particle is certain to be found in some unit cell at any time, regardless of the initial intra-cell position $w$. Therefore the final expression for the amplitude simplifies to 
	\begin{equation}
	\label{eqn:amplitudeResult}
	A(\Delta K)\approx \Big|\int dw \, e^{i\Delta Kw}p(w)\Big|^{2}\,\textrm{.}
	\end{equation}
	
	The result takes the form of an anharmonic Debye-Waller factor \textcolor{blue}{\cite{Maradudin1963PhysRev, Mair1980JPhysC,Lovesey1984Book}}. Equivalently, if we impose periodic boundary conditions on the intra-cell motion, then the motion effectively becomes confined, imparting a static level to the ISF \cite{Jardine2009PCCP}, and the amplitude (\ref{eqn:amplitudeResult}) represents that fictitious static level.
	
	The central approximation (\ref{eqn:factorizationApprox}) neglects the detailed interplay between the spreading of probability from the initial site to neighboring sites, and the randomization of the intra-cell position which occurs in parallel. Additionally, we have not proved that in any given situation the functions $q_{n}(t)$ really do evolve according to a Chudley-Elliott model. However, in any situation where the Chudley-Elliott model is known to provide a good description of the ISF at long times, the result (\ref{eqn:amplitudeResult}) predicts the amplitudes subject to a single approximation represented by the factorization (\ref{eqn:factorizationApprox}) which is based on the separation of timescales between vibrational relaxation and the residence time between jumps.
	
	Regardless of the quality of the factorization approximation, Equation (\ref{eqn:amplitudeResult}) gives the exact long-time limit of the ISF at the diffraction conditions where $\exp(i\Delta Kan')=1$ for all integers $n'$. The factor $\exp(i\Delta Kan')$ in (\ref{eqn:formalISFnwExpansionZeroed}) is constant such that
	\begin{widetext}
		\begin{equation}
		I(\Delta K,t)=\int dw\,dw'\,\sum_{n'} p_{t}(n',w'|0,w)e^{i\Delta Kw'}e^{-i\Delta Kw}p(w)\,\textrm{.}
		\end{equation}
	\end{widetext}
	Performing the internal sum over $n'$ at any $w'$ and $w$ will always yield unity by definition. Then, given that the conditional probability $p_{t}(w'|w)$ at long times relaxes to the unconditional probability $p(w')$, we recover the static level $C(\Delta K)$ of the ISF as the same expression previously derived for the exponential amplitude:
	\begin{equation}
	I(\Delta K,t\rightarrow\infty)\rightarrow C(\Delta K)\,\textrm{;}
	\end{equation}
	\begin{equation}
	\label{eqn:staticExact}
	C(\Delta K)=\Big|\int dw\,p(w)e^{i\Delta Kw}\Big|^{2}\,\textrm{,}
	\end{equation}
	which holds only at diffraction conditions where $e^{i\Delta Ka}=1$. The assumption that $p_{t}(w'|w)\rightarrow p(w')$ at long times depends only on the ergodicity of the Langevin dynamics within one unit cell with periodic boundary conditions, and therefore does not require a separation of timescales between inter-cell and intra-cell motion.
	
	\color{black}
	
	Based on the results (\ref{eqn:amplitudeResult}) and (\ref{eqn:staticExact}), the factorization approximation can be viewed as a physically motivated interpolation of $A(\Delta K)$ between the exact static levels of the ISF at the diffraction conditions of the periodic potential. For weakly corrugated potentials that do not satisfy the required conditions for the approximation (\ref{eqn:factorizationApprox}), the integral (\ref{eqn:amplitudeResult}) over the unit cell still gives the exact static level when $\Delta K$ is a reciprocal lattice vector, but is not guaranteed to accurately predict the amplitude of the slowest decaying component at other values of $\Delta K$. Therefore, for weakly corrugated potentials there is scope for details such as \textcolor{blue}{the strength and frequency dependence of dissipation to influence the relative amplitudes associated with different dynamical processes, as they do in the case of a perfectly flat potential \cite{Townsend2018JPCO}.}
	
	Explicitly, in a sinusoidal potential
	\begin{equation}
	\label{eqn:sinusoidalPotential}
	V(x)=V_{1}\sin\Big(\frac{2\pi x}{a}\Big)\,\textrm{,}
	\end{equation}
	the factorization approximation for the decay amplitude gives us
	\begin{equation}
	\label{eqn:intraWellFactor}
	A(\Delta K)=\Bigg|\frac{\int_{0}^{a}dw\,e^{-\beta V_{1}\sin(2\pi w/a)}e^{i\Delta Kw}}{\int_{0}^{a}dw\,e^{-\beta V_{1}\sin(2\pi w/a)}}\Bigg|^{2}\,\textrm{.}
	\end{equation}
	
	An analytically simpler estimate can be constructed by expanding the probability density $p(w)$ as a Gaussian function centred about the potential well in the unit cell. Such a harmonic approximation predicts a decay amplitude equal to the static level of the harmonic oscillator whose frequency is the frustrated translational frequency of the potential. Then, the amplitude can be expressed as a standard harmonic Debye-Waller factor \cite{Lovesey1984Book}. In Section \ref{sec:numerical}, both the harmonic and the anharmonic estimates are compared to numerical data.
	
	\color{blue}
	\subsection{Interpretation as a Debye-Waller factor}
	\label{subsec:DWfactor}
	
	We have noted that the approximate formula (\ref{eqn:amplitudeResult}) is an anharmonic Debye-Waller (DW) factor, which conventionally describes the reduction in elastic scattered intensity at the Bragg conditions when a scattering probe diffracts from a \textcolor{blue}{solid \cite{Lovesey1984Book} or with caveats a solid surface \cite{Manson1991Scattering}} which may be clean or adsorbate-covered \cite{Gumhalter1999PRB}. The expression (\ref{eqn:staticExact}) for the nonzero static level of the single-particle ISF at the diffraction conditions $e^{i\Delta Ka}=1$, is associated with an elastic contribution to the dynamical structure factor $S(\Delta K,\Delta\omega)$, the Fourier transform of $I(\Delta K,t)$ from the time to the frequency domain. Therefore the form of expression (\ref{eqn:staticExact}) for $C(\Delta K)$ is unsurprising as it can be written as
	\begin{equation}
	C(\Delta K)=\exp[-2W(\Delta K)]\,\textrm{,}
	\end{equation}
	where the implicit definition of the exponent via
	\begin{equation}
	\exp[-W(\Delta K)]=\langle \exp(i\Delta K w)\rangle
	\end{equation}
	matches the standard definition of the Debye-Waller exponent \cite{Lovesey1984Book}.
	
	The DW factor is present even in the absence of adsorbates in which case it arises from the nonzero mean square displacements of substrate atoms. In considering the ISF associated with adsorbate dynamics, we are considering only scattering from adsorbates, not from the substrate. The result (\ref{eqn:staticExact}) shows that the fraction of intensity at the diffraction condition which is scattered elastically, within the kinematic approximation for adsorbate-induced scattering, is attenuated by the DW factor associated with the intra-cell co-ordinate in a periodic adsorbate-substrate potential. The presence of the DW factor in the static level of the ISF for confined vibrational motion has been noted previously \cite{Vega2004JCP,Vega2004JPCM}, and in the present work we connect the idea to the amplitude of the exponential tail of the ISF present when the adsorbates have long range mobility due to hopping motion.
	
	Our approximation (\ref{eqn:amplitudeResult}) for the amplitude of the exponential tail of the ISF is the DW factor $\exp[-2W(\Delta K)]$ evaluated away from the diffraction conditions. In the frequency domain what would be an elastic contribution in the absence of inter-cell hopping, broadens into a Lorentzian with finite width. The expression for the amplitude of the exponential tail in the time domain is proportional to the area of the Lorentzian contribution in the frequency domain. It is well known that the quasi-elastic peak in $S(\Delta K,\Delta\omega)$ cannot be cleanly attributed to jump motion, since pure vibrational line shapes include a quasi-elastic contribution \cite{Vega2004JCP, Vega2004JPCM, Jardine2004JCP, MartinezCasado2010ChemPhys}. In the language of Equation (\ref{eqn:longTimeLimit}), the fast-decaying component that must be added to $A(\Delta K)\exp[-\alpha(\Delta K)t]$ to make up the the full ISF does not have zero mean in general, and therefore contains a quasi-elastic contribution in the frequency domain. However, the simulation results in Section \ref{sec:numerical} show that a component of the ISF associated with jump motion can be cleanly separated out in the time domain. The complementary perspective on surface dynamics offered by the time domain is one reason for the utility of the spin echo method for studying adsorbate jump dynamics \cite{Jardine2009ProgSurfSci}.
	
	\subsection{Information contained in $A(\Delta K)$}
	\label{subsec:infoInAmplitude}
	
	$A(\Delta K)$ is directly dependent on the adsorbate/substrate potential. Within the data reduction $I(\Delta K,t>t^{*})\approx A(\Delta K)\exp[-\alpha(\Delta K)t]$, the decay rates $\alpha(\Delta K)$ are also very strongly sensitive to the dynamical friction $\gamma$, however as shown in the manuscript the amplitudes $A(\Delta K)$ are not. Therefore, the amplitudes provide a measure of $V(w)$, independent of other measures such as $\alpha(\Delta K)$. In the context of modeling surface dynamics as measured by HeSE, predicting the relative amplitudes of different components of multi-component dynamical signatures is an ongoing project. Within a Langevin simulation approach, the potential energy surface $V(\mathbf{R})$ for the adsorbate motion can be tuned to adjust both $\alpha(\Delta K)$ and $A(\Delta K)$. Such an operation has been carried out previously \cite{Ward2013Thesis} using intuition to guide the adjustment of simulation parameters. However, having a simple analytical formula to guide the search would be valuable, and that is what the present work contributes.
	
\begin{figure}
		\centering
		\includegraphics[width=0.45\textwidth]{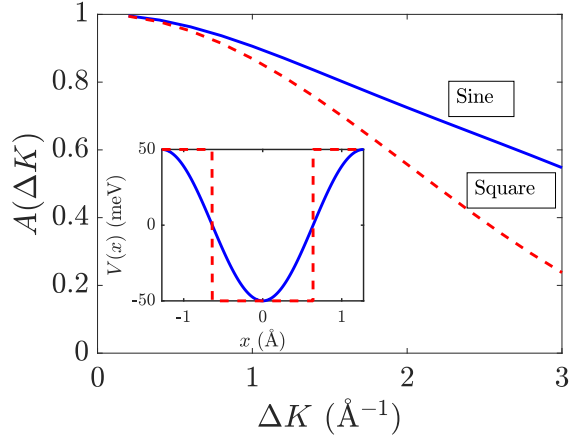}
		\caption{Illustration of the effect of the periodic potential on the amplitudes $A(\Delta K)$, within the approximation encoded by Equation (\ref{eqn:amplitudeResult}). The solid blue curve shows $A(\Delta K)$ evaluated by numerical integration of Equation (\ref{eqn:amplitudeResult}) for the sinusoidal potential $V(x)=V_{1}\sin(2\pi x/a)$. The red dashed curve shows the corresponding resut when the sinusoidal potential is substituted for a square-wave potential with the same barrier height. The potentials are plotted over one unit cell ($a=2.55\,$\AA) in the inset.}
		\label{fig:sinVsSquare}
	\end{figure}

	To illustrate the potentially dramatic effect on $A(\Delta K)$ arising from the shape of the potential energy landscape, in Figure \ref{fig:sinVsSquare} we compare $A(\Delta K)$ from the sinusoidal potential described in Equation (\ref{eqn:intraWellFactor}), with $A(\Delta K)$ from a square-wave potential with the same barrier height.
	
	\color{black}
	
	\subsection{Contribution of surface-perpendicular motion}
	\label{subsec:zmotion}
	
	The factorization approximation for the decay amplitude of the jump diffusion signature should be interpreted here as a way of understanding the amplitudes observed in Langevin simulations. A full connection to the amplitudes observed in experimental measurements that probe the ISF \cite{Jardine2009ProgSurfSci} would require more detailed scattering calculations, as well as consideration of additional inelastic channels which could involve surface phonons and potentially surface-perpendicular vibrations ($z$-motion, or S-mode). Within the simple classical and kinematic approximation underlying Equation (\ref{eqn:harmonicLongTimeLimit}), $z$ motion of typical frequency $\Omega_{z}$ will lead to a reduction in the jump signature decay amplitude by a factor of
	\begin{equation}
	\label{eqn:zmotionLongTimeLimit}
	\exp\Big(-\frac{k_{B}T}{m\Omega_{z}^{2}}\Delta k_{z}^{2}\Big)\,\textrm{.}
	\end{equation}
	
	A more detailed investigation of the intensities of S-mode excitation in the HeSE experimental geometry would be needed to confirm the absolute scale of the effect. However, within the classical kinematic estimate (\ref{eqn:zmotionLongTimeLimit}) the influence of $z$-motion is  most pronounced for adsorbates with small $m\Omega_{z}^{2}$. Combined with the large typical surface-perpendicular momentum transfer $\Delta k_{z}\approx 6\,$\AA$^{-1}$ on the Cambridge spin echo spectrometer with its $44.4^{\circ}$ fixed scattering angle \cite{Fouquet2005RSI} and $8\,$meV standard beam energy, a large reduction in the measured decay amplitude due to $z$-motion is highly plausible unless the S-mode excitation is strongly suppressed. Well-known model adsorption systems exhibiting relatively low $m\Omega_{z}^{2}$ include Na/Cu(001) and several Xe/metal systems \cite{Graham2003SurfSciRep}. For Xe/Pt(111), at a surface temperature $T=121\,$K and using $\hbar\Omega_{z}=3.6\,$meV \cite{Bruch2000JCP} and $\Delta k_{z}=6\,$\AA$^{-1}$, Equation (\ref{eqn:zmotionLongTimeLimit}) predicts a reduction of the jump signature amplitude by a factor of $0.5$. Therefore although an intriguing ``table-top" model potential for Xe/Pt(111), within a two-dimensional Langevin framework, was successful in explaining the experimental relative amplitudes of ballistic and jump-motion components \cite{Ward2013Thesis}, the surface-perpendicular degree of freedom is highly relevant to any attempt to explain the absolute amplitudes in the ISF for xenon systems.
	
	\section{Numerical results}
	\label{sec:numerical}
	
	\subsection{One spatial dimension}
	\label{sec:oneSpatialDimension}
	
\begin{figure}
		\centering
		\includegraphics[width=0.45\textwidth]{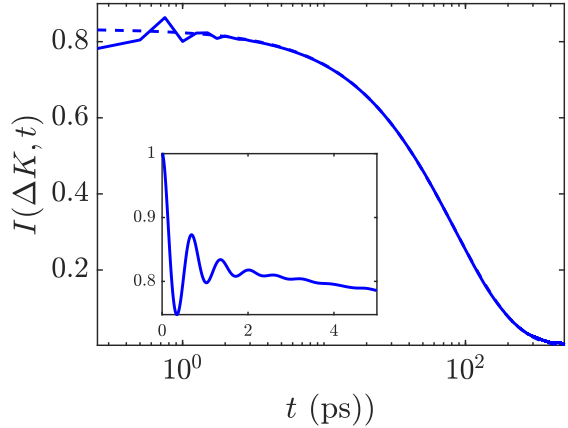}
		\caption{Numerical ISF at $\Delta K=1.6\,$\AA$^{-1}$, computed from one-dimensional Langevin simulations performed performed for a particle of mass $m=28\,$amu  in the sinusoidal model potential of Equation (\ref{eqn:sinusoidalPotential}) with $V_{1}=50\,$meV, $a=2.55\,\textrm{\AA}$, at $T=200\,$K and $\gamma=3\,$ps$^{-1}$ using a timestep of $5\,$fs and Verlet integration \cite{Verlet1967PhysRev}, for $300$ independent runs. The blue solid line represents the simulation data, and the blue dashed line shows the best mono-exponential fit to the long time tail. The intercept of the fitted curve extrapolated back to $t=0$ is equivalent to the decay amplitude. In the main plot, the ISF is sampled every $0.25\,$ps. In the inset, the short time behaviour of the ISF is shown at a finer time resolution.}
		\label{fig:langevin1dISF}
	\end{figure}

\begin{figure}
		\includegraphics[width=0.45\textwidth]{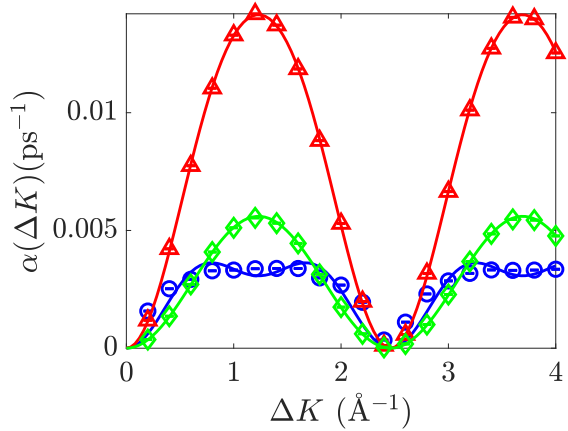}
		\caption{The $\Delta K$-dependent decay rate $\alpha(\Delta K)$, obtained by fitting the tail of the ISF to the functional form (\ref{eqn:longTimeLimit}). Blue circles, red triangles and green diamonds respectively are the results of simulations at $\gamma=0.3$, $3.0$ and $30\,$ps$^{-1}$ respectively. $\alpha(\Delta K)$ is sinusoidal in $\Delta K$ and returns to a perfect minimum at the diffraction condition $\Delta K a=2\pi$, indicating jump motion. The solid curves is the best fit to the one-dimensional Chudley-Elliott model (\ref{eqn:ceModel1d}) allowing for jumps up to third nearest neighbour in length. The absolute scale of $\alpha(\Delta K)$ is strongly dependent on the friction, in agreement with the expectations of classical rate theory \cite{Hanggi1990RevModPhys}\textcolor{blue}{\cite{Guantes2003JCP}}. By contrast, the amplitudes in Figure \ref{fig:langevin1dAmplitude} display almost no dependence on the friction. \textcolor{blue}{We emphasize that the $\alpha(\Delta K)$ and associated jump rates, in the present work, are not computed independently using analytical rate theories, but are extracted freely from the Langevin simulation data. Although analytical methods can be applied to compute accurate rates in two-dimensional diffusion problems \cite{Guantes2003JCP}, any approximation introduced by constraining $\alpha(\Delta K)$ will bias the extracted values of $A(\Delta K)$, and therefore it is desirable to fit $A(\Delta K)$ and $\alpha(\Delta K)$, as a free and mutually consistent pair of fitting parameters, from the numerical simulated ISFs.}}
		\label{fig:langevin1dAlphaDeltaK}
	\end{figure}

\begin{figure}
		\centering
		\includegraphics[width=0.45\textwidth]{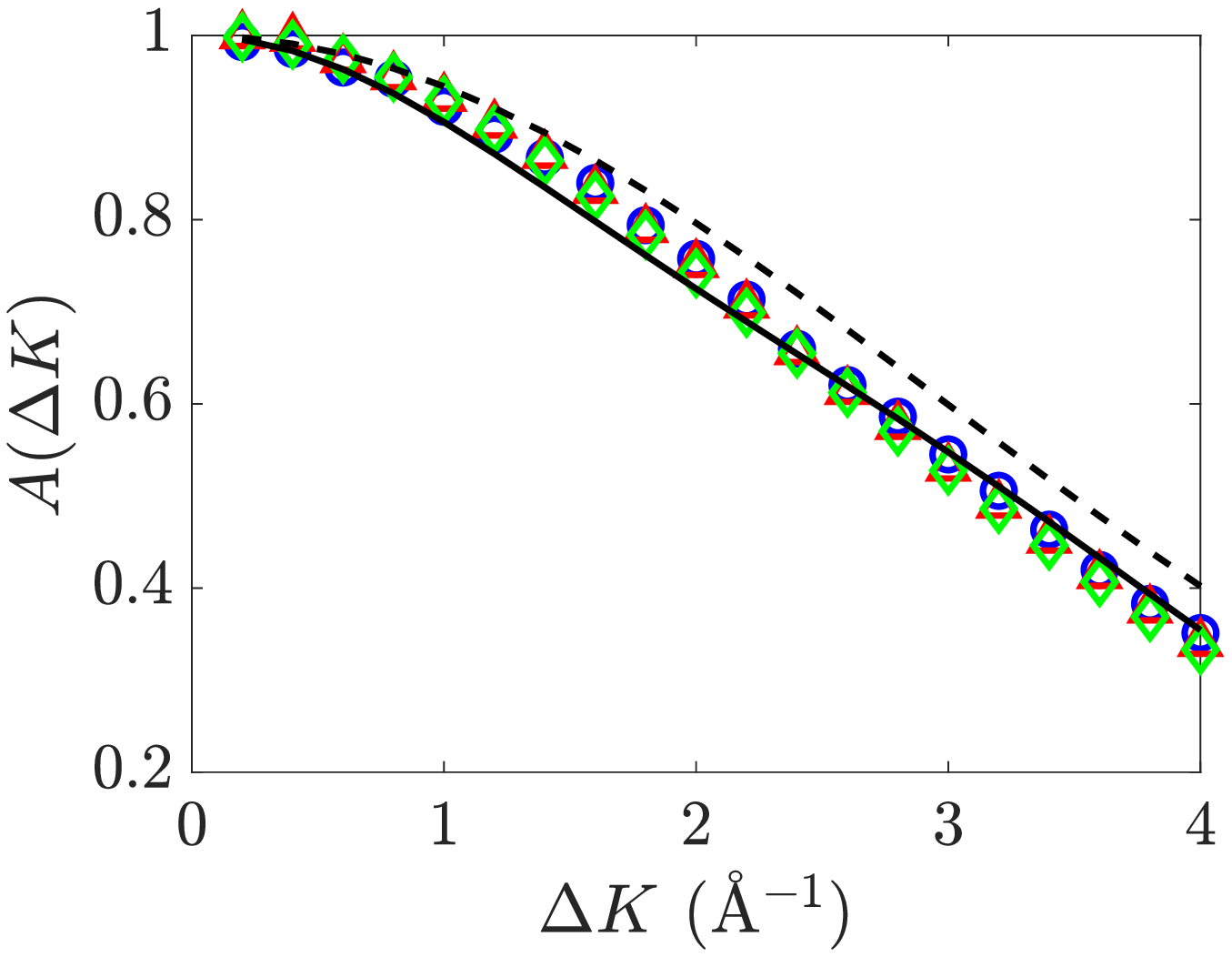}
		\caption{The $\Delta K$-dependent decay amplitudes $A(\Delta K)$, associated with the same ISFs whose decay rates are presented in Figure \ref{fig:langevin1dAlphaDeltaK}. Unlike the $\alpha(\Delta K)$, the amplitude $A(\Delta K)$ is not sinusoidal and does not return to zero at the diffraction condition, as the amplitude is related to motion in the continuous potential, not on a discrete lattice. Additionally, the amplitudes in Figure \ref{fig:langevin1dAmplitude} display almost no dependence on the friction strength, in contrast to the $\alpha(\Delta K)$. The black solid curve overlaid is the prediction of Equation (\ref{eqn:intraWellFactor}) numerically integrated over one unit cell of $V(x)$. It is not a fit to the simulation data. The simple estimate based on the harmonic well frequency (Equation \ref{eqn:harmonicLongTimeLimit}) is plotted as the black dashed curve.}
		\label{fig:langevin1dAmplitude}
	\end{figure}

	The analytical estimates for the decay amplitude can be tested by numerical Langevin simulations. The Langevin simulations whose results are illustrated in Figures \ref{fig:langevin1dISF}-\ref{fig:langevin1dAmplitude} were performed for a particle of mass $m=28\,$amu  in the sinusoidal model potential of Equation (\ref{eqn:sinusoidalPotential}) with $V_{1}=50\,$meV, $a=2.55\,\textrm{\AA}$, at $T=200\,$K and $\gamma=\{0.3,3,30\}\,$ps$^{-1}$ using a timestep of $5\,$fs and Verlet integration \cite{Verlet1967PhysRev}. $300$ independent simulation runs were performed. The ISF is calculated from each simulated trajectory $x(t)$ using the Wiener-Khinchin theorem in the standard way \cite{Rittmeyer2016PRL,Townsend2018JPCO}, and the ISFs from the separate runs are combined incoherently. Therefore, the average in expression (\ref{eqn:genericClassicalISF}) is partly performed by time averaging, and partly by ensemble averaging. The long time tail of the ISF is fitted by nonlinear least squares from a start time later than $1/\gamma$.
	
	Figure \ref{fig:langevin1dISF} shows a numerical ISF at $\Delta K=1.6\,$\AA$^{-1}$, from the $\gamma=3\,$ps$^{-1}$ simulation. The blue solid curve represents the simulation data, and the blue dashed curve shows the best mono-exponential fit to the long time tail. The two curves overlay each other for the majority of the plot. The intercept of the fitted curve extrapolated back to $t=0$ is equivalent to the decay amplitude. In the main plot, the ISF is sampled every $0.25\,$ps. In the inset, the short time behavior of the ISF is shown at a finer time resolution.
	
	Figure \ref{fig:langevin1dAlphaDeltaK} shows the dephasing rates $\alpha(\Delta K)$ extracted from the simulations at the three different values of the friction. Results for $\gamma=0.3$, $3.0$ and $30\,$ps$^{-1}$ are shown as blue circles, red triangles and green diamonds respectively. The solid curves are fits to the Chudley-Elliott model in one dimension as specified by Equation (\ref{eqn:ceModel1d}). Jumps up to third nearest neighbour in length are included in the fit. The minor disagreement between the fit and the simulated data at low friction is likely due to the presence of additional jumps. However, the important feature of the $\alpha(\Delta K)$ is the unambiguous zero at the diffraction condition, showing that the fitted component of the ISF describes jump motion.
	
	\textcolor{blue}{Figure \ref{fig:langevin1dAmplitude} shows the fitted decay amplitudes associated with the same simulations whose $\alpha(\Delta K)$ were fitted and plotted in Figure \ref{fig:langevin1dAlphaDeltaK}. Unlike the $\alpha(\Delta K)$ which is very strongly dependent on the dissipation strength $\gamma$, the amplitudes are barely affected by $\gamma$ over the two orders of magnitude considered. The factorization approximation gives an accurate estimate of the decay amplitude, with or without the simplifying harmonic approximation, for the conditions simulated.} There are small systematic errors between the theory and the numerical amplitudes, and the errors change sign about the first order diffraction condition. Formulating the decay amplitudes in terms of the real space correlation $G(x,t)$ could give further insight into the required corrections to the factorization approximation. However, the analytical approximations as they stand can clearly be used to understand the trends and the absolute scale of the amplitudes fitted from simulations.
	
	\subsection{Two spatial dimensions}
	\label{sec:twoDimensions}
	
	In two dimensions, we should expect the expression (\ref{eqn:intraWellFactor}) to generalize straightforwardly to a two-dimensional integral over a unit cell $S$,
	\begin{equation}
	\label{eqn:intraWellFactor2d}
	A(\Delta\mathbf{K})=\Bigg|\frac{\int_{S} d\mathbf{R}\,e^{-\beta V(\mathbf{R})}e^{i\Delta \mathbf{K}\cdot\mathbf{R}}}{\int_{S} d\mathbf{R}\,e^{-\beta V(\mathbf{R})}}\Bigg|^{2}\,\textrm{.}
	\end{equation}
	
	The two dimensional result can be verified for a non-separable potential $V(\mathbf{R})$ in two spatial dimensions by simulating
	\begin{equation}
	\label{eqn:langevin}
	m\ddot{\mathbf{R}}=-\mathbf{\nabla}V(\mathbf{R})-m\gamma\dot{\mathbf{R}}+\mathbf{F}(t) \, ,
	\end{equation}
	where isotropic friction will be assumed, and $\langle F_{i}(t) F_{j}(0)\rangle=2mk_{B}T\delta(t)\delta_{i,j}$ with $\delta(t)$ the Dirac delta function and $\delta_{i,j}$ a Kronecker-$\delta$ symbol over the two Cartesian spatial directions of the simulation. 
	
	The potential energy surface is taken as a low order Fourier expansion with the symmetry of a hexagonal close packed surface \cite{Alexandrowicz2008JACS,Hedgeland2011PRL,Rittmeyer2016PRL}. Let $\{ \mathbf{G} \}$ be three linearly independent first-order reciprocal lattice vectors of the surface, whose close packing distance in real space is $a$. Then, a simple Fourier series representation with the correct symmetry for the fcc(111) surface is given by \cite{Alexandrowicz2008JACS,Lechner2013JCP}
	\begin{equation}
	\label{eqn:degeneratePotential}
	V(\mathbf{R})=V_{2}\sum_{\mathbf{G}}\cos(\mathbf{G}\cdot\mathbf{R})\,\textrm{.}
	\end{equation}
	Explicitly, in Cartesian co-ordinates where the primitive surface unit cell in real space is oriented with an axis along the $\mathbf{x}$ direction, the potential is given by
	\begin{widetext}
		\begin{equation}
		V(x,y)=A\Big\{ \cos\Big(\frac{4\pi y}{a\sqrt{3}}\Big)+\cos\Big(\frac{2\pi}{a}\Big[x-\frac{y}{\sqrt{3}} \Big] \Big)+\cos\Big(\frac{2\pi}{a}\Big[x+\frac{y}{\sqrt{3}} \Big] \Big)\Big\}\, \textrm{.}
		\end{equation}
	\end{widetext}
	If $V_{2}<0$ then adsorption occurs at top sites \cite{Alexandrowicz2008JACS}, so we should expect Chudley-Elliott behaviour in the jump dynamics. The saddle points of the potential, providing the minimum energy path for surface diffusion, are the twofold bridge sites which present an energy barrier of $4V_{2}$. We use $V_{2}=25\,$meV to reproduce the same diffusion barrier as in the one dimensional simulations.
	
\begin{figure}
		\includegraphics[width=0.45\textwidth]{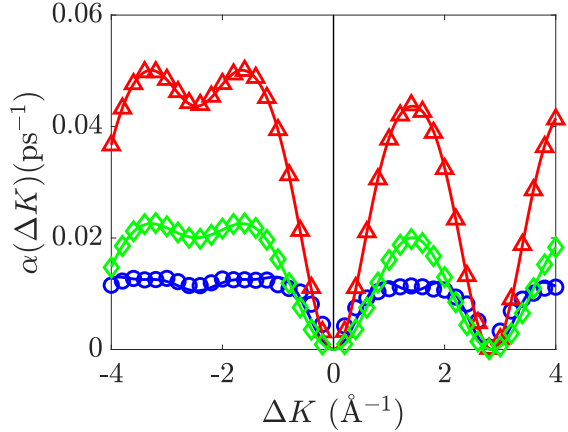}
		\caption{The $\Delta K$-dependent decay rate $\alpha(\Delta K)$ of ISFs at different $\Delta K$. The ISFs were obtained from Langevin simulations in the model potential (\ref{eqn:degeneratePotential}), at the same simulation temperature ($T=200\,$K) and frictions ($\gamma=0.3,3,30\,$ps$^{-1}$) as used in the one-dimensional simulations in Section \ref{sec:oneSpatialDimension}. The negative (positive) $\Delta K$ on the plot corresponds to $\Delta \mathbf{K}$ projected along the $x$ ($y$) direction in the simulation. $\alpha(\Delta K)$ is sinusoidal in $\Delta K$ and returns to a perfect minimum at the diffraction condition $\Delta K a=2\pi$, indicating jump motion. The solid curves are fits to the Chudley-Elliott model (\ref{eqn:ceModel}) for jump diffusion on a two dimensional Bravais lattice, including jumps up to third nearest neighbour in length. As in the case of one-dimensional motion, the absolute scale of the $\alpha(\Delta K)$ is strongly sensitive to the friction strength, whereas the decay amplitudes shown in Figure \ref{fig:langevin2dAmplitude} are not.}
		\label{fig:langevin2dAlphaDeltaK}
	\end{figure}

	Figure \ref{fig:langevin2dAlphaDeltaK} shows the dephasing rates $\alpha(\Delta K)$ extracted from simulations of the two-dimensional LE (\ref{eqn:langevin}) with the same three values of the friction explored in the one dimensional case, and at the same simulation temperature. The momentum transfer $\Delta K$ is projected along the $x$ direction, shown as negative $\Delta K$, and the $y$ direction shown as positive $\Delta K$. Results for $\gamma=0.3$, $3$ and $30\,$ps$^{-1}$ are shown as blue circles, red triangles and green diamonds respectively. The solid curves are fits to the Chudley-Elliott model in two dimensions for jump motion on a Bravais lattice of adsorption sites, allowed jump vectors $\mathbf{j}$ and corresponding rates $\Gamma_{j}$:
	\begin{equation}
	\label{eqn:ceModel}
	\alpha(\Delta\mathbf{K})=2\sum_{\mathbf{j}}\Gamma_{j}\sin^{2}\Big(\frac{\Delta\mathbf{K}\cdot \mathbf{j}}{2}\Big) \textrm{.}
	\end{equation}
	Jumps up to third nearest neighbour in length are included in the fit. The minor disagreement between the fit and the simulated data at low friction is likely due to the presence of additional jumps. However, the important feature of the $\alpha(\Delta K)$ is the unambiguous zero at the diffraction condition, showing that the fitted component of the ISF describes jump motion.
	
\begin{figure}
		\includegraphics[width=0.45\textwidth]{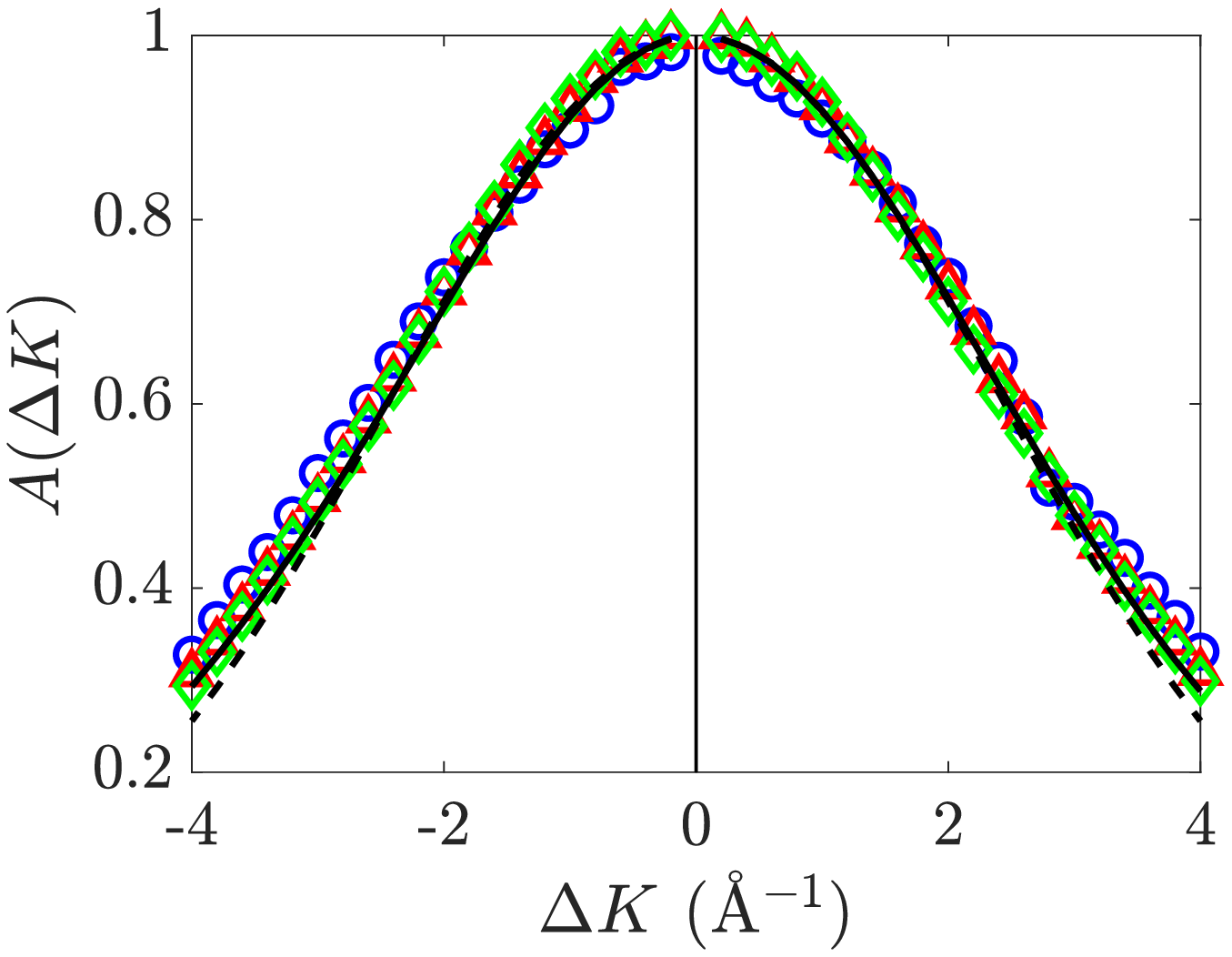}
		\caption{The $\Delta K$-dependent decay amplitudes $A(\Delta K)$, associated with the same ISFs whose decay rates are presented in Figure \ref{fig:langevin2dAlphaDeltaK}. As for the one-dimensional results, the decay amplitude decreases monotonically with $\Delta K$, and is almost independent of the friction strength. Further, $A(\Delta K)$ is approximately isotropic, which reflects the local isotropy of the potential about its minima. The dashed black curve is the prediction using the harmonic approximation via Equation (\ref{eqn:harmonicLongTimeLimit}). The solid black curve is the decay amplitude given by a numerical integration of Equation (\ref{eqn:intraWellFactor2d}) to evaluate the anharmonic Debye-Waller factor. The harmonic and anharmonic estimates both agree closely with the fitted simulation data.}
		\label{fig:langevin2dAmplitude}
	\end{figure}

	Figure 	\ref{fig:langevin2dAmplitude} shows the decay amplitudes that correspond to the decay rates in Figure \ref{fig:langevin2dAlphaDeltaK}. Unlike the $\alpha(\Delta\mathbf{K})$, the amplitudes are almost isotropic, which reflects the isotropic harmonic expansion of the potential at the minima. The amplitudes decay monotonically with $\Delta K$, and are in good agreement with either a harmonic approximation (dashed black curve) or a numerical integration of Equation (\ref{eqn:intraWellFactor2d}) (solid black curve).
	
	\section{Conclusions}
	\label{sec:conc}
	
	The dynamics of a classical particle, obeying the Langevin in a periodic potential with a Bravais lattice arrangement of well defined adsorption minima, have been studied via the intermediate scattering function (ISF), the autocorrelation of the kinematic scattering amplitude. Under all conditions studied, the long time limit of the ISF is an exponential decay whose decay rate is in excellent agreement with an idealized jump model, confirming that the long time dynamics are well described as jump motion. The amplitude of the jump diffusion signature is smaller than unity, due to the intra-cell dynamics. \textcolor{blue}{The present work provides} an analytical approximation by which to quantitatively understand the momentum transfer dependence of the jump signature amplitude, for strongly corrugated systems exhibiting a clean separation between jump and intra-cell motion.
	
	\textcolor{blue}{The amplitude of the jump component is accurately described by a factorization approximation, in which the conditional probability density of the particle co-ordinate at long times is assumed to relax into the product of an intra-cell term and a probability to be found in different unit cells. The factorization approximation predicts a simple thermodynamic integral expression for the amplitude of the long-time, exponential decay limit of the ISF describing jump motion. The factorization approximation is exact when the ISF is evaluated at a momentum transfer corresponding to a reciprocal lattice of the periodic potential, at which points the formula comprises a classical anharmonic Debye-Waller factor for intra-cell motion. Comparison with Langevin molecular dynamics simulations in strongly corrugated potentials shows that the factorization formula also provides an accurate interpolation of the simulated amplitudes between the diffraction conditions.}
	
	\begin{acknowledgments}
		PT thanks the UK Engineering and Physical Sciences Research Council for doctoral funding under award number 1363145. The authors thank Dr Bill Allison for helpful discussions.
	\end{acknowledgments}

\end{document}